\documentclass[preprint,amsmath,amssymb,aps,notitlepage]{revtex4-1}

\usepackage{graphicx}
\usepackage{gensymb}
\usepackage[sort&compress]{natbib}
\usepackage[symbol]{footmisc}
\usepackage{caption}
\usepackage{setspace}
\usepackage{subcaption}
\usepackage{fullpage}
\usepackage{multirow}
\usepackage{fancyhdr}
\makeatletter
\renewcommand\@make@capt@title[2]{%
 \@ifx@empty\float@link{\@firstofone}{\expandafter\href\expandafter{\float@link}}%
  {\textbf{#1}}\@caption@fignum@sep#2\quad
}%
\makeatother

\begin{document}

\title{Teaching critical thinking}

\author{N.G. Holmes}
\email[]{ngholmes@stanford.edu}
\affiliation{Department of Physics, Stanford University, Stanford, CA}
\author{Carl E. Wieman}
\affiliation{Department of Physics, Stanford University, Stanford, CA}
\affiliation{Graduate School of Education, Stanford University, Stanford, CA}
\author{D.A. Bonn}
\affiliation{Department of Physics and Astronomy, University of British Columbia, Vancouver, BC}

\begin{abstract}
{The ability to make decisions based on data, with its inherent uncertainties and variability, is a complex and vital skill in the modern world. The need for such quantitative critical thinking occurs in many different contexts, and while it is an important goal of education, that goal is seldom being achieved. We argue that the key element for developing this ability is repeated practice in making decisions based on data, with feedback on those decisions. We demonstrate a structure for providing suitable practice that can be applied in any instructional setting that involves the acquisition of data and relating that data to scientific models. This study reports the results of applying that structure in an introductory physics lab course. Students in an experimental condition were repeatedly instructed to make and act on quantitative comparisons between datasets, and between data and models, an approach that is common to all science disciplines. These instructions were slowly faded across the course. After the instructions had been removed, students in the experimental condition were 12 times more likely to spontaneously propose or make changes to improve their experimental methods than a control group, who performed traditional experimental activities. They were also four times more likely to identify and explain a limitation of a physical model using their data. Students in the experimental condition also showed much more sophisticated reasoning about their data. These differences between the groups were seen to persist into a subsequent course taken the following year.}

\bigskip
\noindent\emph{Significance Statement}: Understanding and thinking critically about scientific evidence is a crucial skill in the modern world. We present a simple learning framework that employs cycles of decisions about making and acting on quantitative comparisons between datasets or data and models. With opportunities to improve the data or models, this structure is appropriate for use in any data-driven science learning setting. This structure led to significant and sustained improvement in students' critical thinking behaviours, compared to a control group, with effects far beyond that of statistical significance.

\bigskip
\noindent\emph{Keywords}: critical thinking | scientific reasoning | scientific teaching | teaching experimentation | undergraduate education
\end{abstract}

\maketitle


A central goal of science education is to teach students to think critically about scientific data and models. It is crucial for scientists, engineers, and citizens in all walks of life to be able to critique data, to identify whether or not conclusions are supported by evidence, and to distinguish a significant effect from random noise and variability. There are many indications of how difficult it is for people to master this type of thinking as evidenced by many societal debates. Although teaching quantitative critical thinking is a fundamental goal of science education, particularly the laboratory portion, the evidence indicates this is seldom, if ever, being achieved \cite{KanariMillar, Kumassah, KungLinder, RyderLeach,RyderRight,Sere1993}. To address this educational need, we have analyzed the explicit cognitive processes involved in such critical thinking and then developed an instructional design to incorporate these processes. 

We argue that scientists engage in such critical thinking through a process of repeated comparisons and decisions: comparing new data to existing data and/or models, and then deciding how to act on those comparisons based on analysis tools that embody appropriate statistical tests. Those actions typically lead to further iterations involving improving the data and/or modifying the experiment or model. In a research setting, common decisions are to improve the quality of measurements (in terms of accuracy or precision) to determine whether an effect is hidden by large variability, to embrace, adjust or discard a model based on the scientific evidence, or to devise a new experiment to answer the question. In other settings, such as medical policy decisions, there may be fewer options but corresponding decisions are made as to the consistency of the model and the data and what conclusions are justified by the data. 

We hypothesize that much of the reason students do not engage in these behaviours is because the educational environment provides few opportunities for this process. Students ought to be explicitly exposed to how experts engage in critical thinking in each specific discipline, which should, in turn, expose them to the nature of knowledge in that discipline \cite{Baron1993}. Demonstrating the critical thinking process, of course, is insufficient for students to use it on their own. Students need practice engaging in the critical thinking process themselves, and this practice should be deliberate and repeated with targeted feedback \cite{Baron1993,DeliberatePractice, Kuhn2008}. We do not expect first year university students to engage in expert-level thinking processes. We can train them to think more like scientists by simplifying the expert decision tree described above. Making it explicit to students, demonstrating how it allows them to learn or make discoveries, and having them practice in a deliberate way with targeted feedback, will help students understand the nature of scientific measurement and data uncertainty, and, in time, adopt the new ways of thinking.

The decision tree and iterative process we have described could be provided in any setting in which data and models are introduced to students. Virtually all instructional labs in science offer such opportunities as students collect data and use it to explore various models and systems. Such labs are an ideal environment for developing students' critical thinking and this is arguably their greatest value beyond simply skills-training. 

We have tested this instructional concept in the context of a calculus-based introductory laboratory course in physics at a research-intensive university. The students repeatedly and explicitly make decisions and act on comparisons between data sets or between data and models as they work through a series of simple, introductory physics experiments. Although this study is in the context of a physics course, we believe the effect would be similar using experiments from any subject that involve quantitative data, opportunities to quantitatively compare data and models, and opportunities to improve data and models. With this simple intervention, we observed dramatic long-term improvements in students' quantitative critical thinking behaviours when compared with a control group that carried out the same lab experiments, but with a structure more typical of instructional labs.  

\begin{figure}[h]
\includegraphics[width=0.4\textwidth]{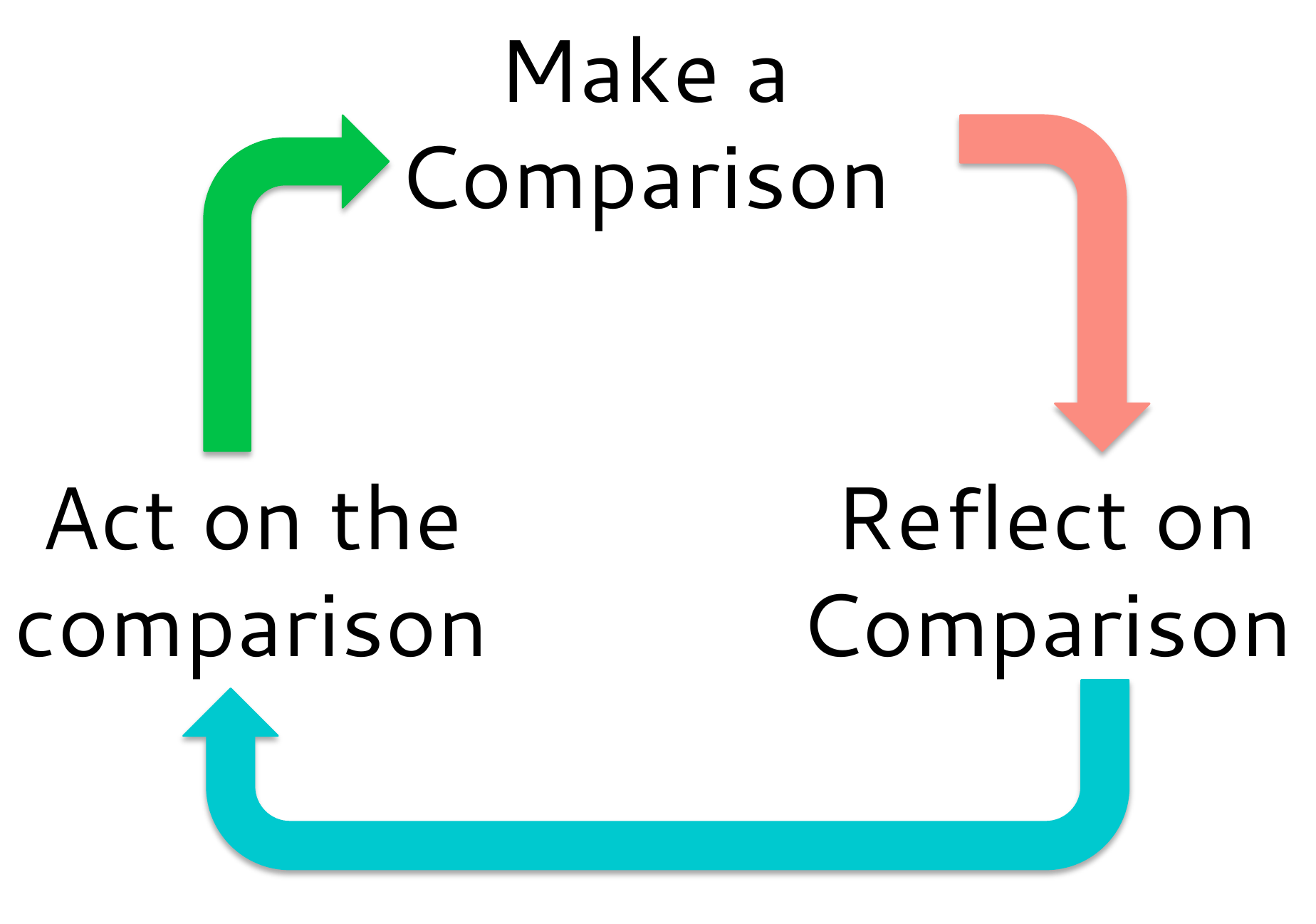}
\caption{The experimental condition engaged students in iterative cycles of making and acting on comparisons of their data. This involved comparing pairs of measurements with uncertainty or comparing data sets to models using weighted $\chi^2$ and residual plots.}\label{fig:cycle}
\end{figure}

In our study, students in the experiment condition were explicitly instructed to (and received grades to) quantitatively compare multiple collected data sets or a collected data set and a model, and to decide how to act on the comparisons (figure \ref{fig:cycle}). While a variety of options for acting on comparisons, as listed above, were presented to students, striving to improve the quality of their data was the most rigorously enforced. For example, in one of the earliest experiments, students were told to make two sets of measurements and compare them quantitatively. They were then prompted to devise a plan to improve the quality of their measurements, to discuss this plan with other groups, and to carry out the revised measurements and analysis. This explicit focus on measurements, rather than improving models, was intended to address the fact that students in a lab course often assume data they collect is inherently low quality compared to expert results \cite{Allie1998}. This can lead them to ignore disagreements between measurements or to artificially inflate uncertainties to disguise them \cite{PERC}. When disagreements do arise, students often attribute them to what they refer to as `human error' \cite{Sere2001} or simply blame the equipment being used. As such, students are unlikely to adjust or discard an authoritative model, since they do not trust that their data is sufficiently high quality to make such a claim. We hypothesize that the focus on high-quality data will, over time, encourage students to critique models without explicit support. 

To compare measurements quantitatively, students were taught a number of analysis tools used regularly by scientists in any field. Students were also taught a framework for how to use these tools to make decisions about how to act on the comparisons. For example, they were shown weighted $\chi^2$ calculations for least squares fitting of data to models and then were given a decision tree for interpreting the outcome. If they obtain a low $\chi^2$ they would decide whether it means their data is in good agreement with the model or whether it means they have overestimated their uncertainties. If they obtain a large $\chi^2$ they would decide whether there is an issue with the model or with the data. From these interpretations, the decision tree expands into deciding what to do. In both cases, students were encouraged to improve their data: to improve precision and decrease their uncertainties in the case of low $\chi^2$, or to identify measurement or systematic errors in the case of a large $\chi^2$. While they were told that a large $\chi^2$ might reflect an issue with the model, they were not told what to do about it, leaving room for autonomous decision making. Regardless of the outcome of the comparison, therefore, students had guidelines for how to act on the comparison, typically leading to additional measurements. This naturally led to iterative cycles of making and acting on comparisons, which could be used for any type of comparison. 

Before working with $\chi^2$ fitting and models, students were first introduced to an index for comparing pairs of measured values with uncertainty (the ratio of the difference between two measured values to the uncertainty in the difference, see S1.1 for more details). Students were also taught to plot residuals (the point-by-point difference between measured data and a model) to visualize the comparison of data and models. Both of these tools, and any comparison tool that includes the variability in a measurement, lend themselves to the same decision process as the $\chi^2$ value when identifying disagreements with models or improving data quality. A number of standard procedural tools for determining uncertainty in measurements or fit parameters were also taught (see S1.1 for the full list). As more tools were introduced during the course, the explicit instructions to make or act on the comparisons were faded (see S1.2 for more details and for a week-by-week diagram of the fading).

The students carried out different experiments each week and completed the analysis within the three-hour lab period. To evaluate the impact of the comparison cycles, we assessed students' written lab work from three lab sessions (see S1.3 for a description of the experiments) from the course: one early in the course when the experimental group had explicit instructions to perform comparison cycles to improve data (week 2), and two when all instruction about making and acting on comparisons had been stopped (weeks 16 and 17). We also examined student work from a quite different lab course taken by the same students in the following year. About a third of the students from the first-year lab course progressed into the second year (sophomore) physics lab course. This course had different instructors, experiments, and structure. Students carried out a smaller number of more complex experiments, each one completed over two weeks, with final reports then submitted electronically. We analyzed the student work on the third experiment in this course.  

\section*{Results}
Students' written work was evaluated for evidence of acting on comparisons, either suggesting or executing changes to measurement procedures, or critiquing or modifying physical models in light of collected data. We also examined students' reasoning about data to further inform the results (see S1.4 for inter-rater reliability of the coding process for these three measures). Student performance in the experimental group ($n\approx130$) was compared with a control group ($n\approx130$). The control was a group of students who had taken the course the previous year with the same set of experiments. Analysis in the supplementary material demonstrates that the groups were equivalent in performance on conceptual physics diagnostic tests (S1.5). Although both groups were taught similar data analysis methods (such as weighted $\chi^2$ fitting), the control group was neither instructed nor graded on making or acting on cycles of quantitative comparisons. They also were not introduced to plotting residuals or comparing differences of pairs of measurements as a ratio of the combined uncertainty. However, instructions given to the experimental group were faded over time, so the instructions given to both groups were identical in week 16 and week 17. 

\begin{figure*}[ht]
\includegraphics[width=\textwidth]{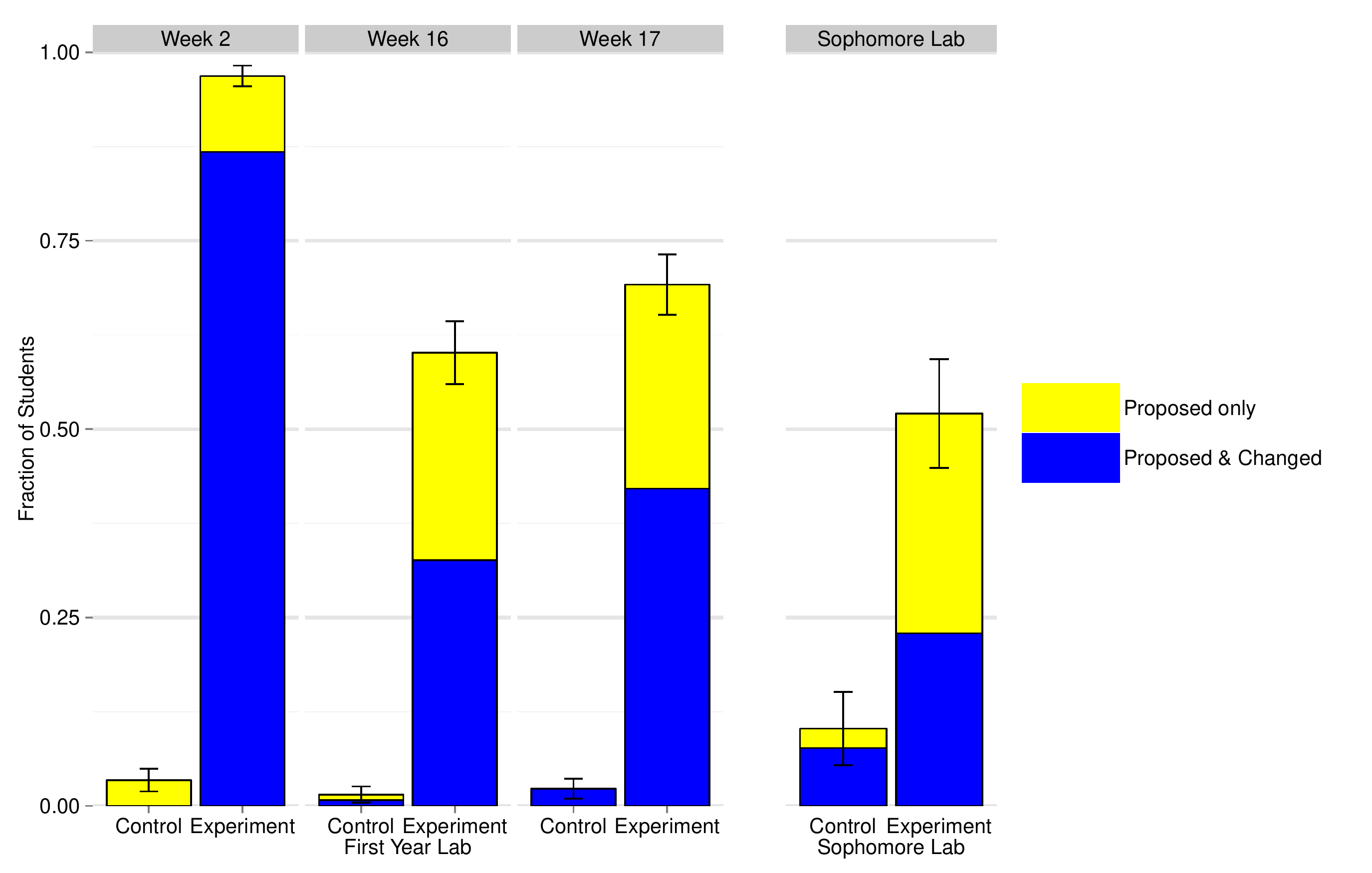}
\caption{\textbf{Method Changes.} The fraction of students proposing and/or carrying out changes to their experimental methods over time shows a large and sustained difference between the experimental and control groups. This difference is substantial when students in the experimental group were prompted to make changes (Week 2), but continues even when instructions to act on the comparisons are removed (Week 16 and 17). This even occurs into the sophomore lab course (see S2.1 for statistical analyses). Note that the sophomore lab data represents a fraction (~1/3) of the first year lab population. Uncertainty bars represent 67\% confidence intervals on the total proportions of students proposing or carrying out changes in each group each week.}\label{fig:iterate}
\end{figure*}

We first compiled all instances where students decided to act on comparisons by proposing and/or making changes to their methods (figure \ref{fig:iterate}), since this was the most explicitly structured behaviour for the experimental group. When students in the experimental group were instructed to iterate and improve their measurements (week 2), nearly all students proposed or carried out such changes. By the end of the course, when the instructions had been removed, over half of the experimental group continued to make or propose changes to their data or methods. This fraction was similar for the sophomore lab experiment, where it was evident that they were making changes, even though we were evaluating final reports rather than laboratory notebooks. Almost none of the control group did at any time.

\begin{figure*}[ht]
\begin{center}
\includegraphics[width=\textwidth]{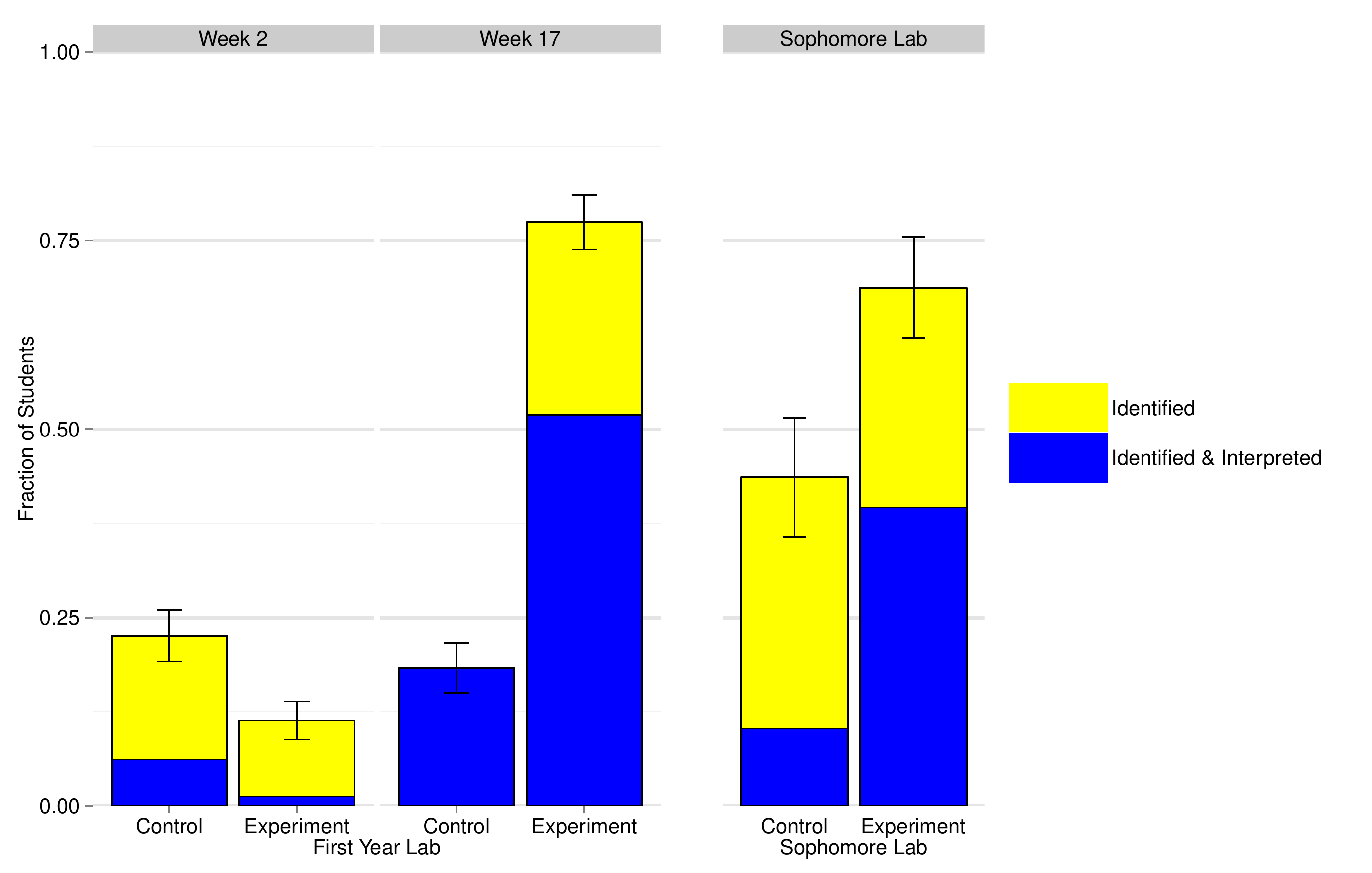}
\caption{\textbf{Evaluating Models.} The fraction of students that identified and correctly interpreted disagreements between their data and a physical model shows significant gains by the experiment group across the lab course (see S2.2 for statistical analyses). This effect is sustained into the sophomore lab. Note that the sophomore lab students were prompted about an issue with the model, which explains the increase in the number of students identifying the issue in the control group. Uncertainty bars represent 67\% confidence intervals on the total proportions of students identifying or interpreting the model disagreements in each group each week.}\label{fig:eval}
\end{center}
\end{figure*}

Next, we looked for instances where students decided to act on a comparison by critiquing the validity of a given physical model (figure \ref{fig:eval}). For both groups of students, many experiments asked them to verify the validity of a physical model. Neither group, however, received explicit prompts to identify or explain a disagreement with the model. Three experiments (week 2, week 17, and the sophomore lab) were included in this portion of the analysis, since these experiments involved physical models that were limited or insufficient for the quality of data achievable (see S1.3). In all three experiments, students' written work was coded for whether they identified a disagreement between their data and the model and whether they correctly interpreted the disagreement in terms of the limitations of the model.  

As shown in figure \ref{fig:eval}, few students in either group noted a disagreement in week 2. As previously observed, learners tend to defer to authoritative information \cite{Baron1993, Allie1998, PERC}. In fact, many students in the experimental group stated that they wanted to improve their data to get better agreement, ignoring the possibility that there could be something wrong with the model. 

As they progress in the course, however, dramatic changes emerge. In week 17, over 3/4 of the students in the experimental group identified the disagreement, nearly four times more than in the control group, and over half of the experimental group provided the correct physical interpretation. Students in the experimental group showed similar performance in the sophomore lab, indicating that the quantitative critical thinking was carried forward. The lab instructions for the sophomore experiment provided students with a hint that a technical modification to the model equation may be necessary if the fit was unsatisfactory, and prompted them to explain why it might be necessary. This is probably why a larger percentage of students in the control group identified the disagreement in this experiment than in the week 2 and 17 experiments. However, only 10\% of the students in the control group provided the physical interpretation, compared to 40\% in the experiment group.

The more sophisticated analysis of models depends on the repeated attempts to improve the quality of the measurements. Students obtain both better data and greater confidence in the quality of their data, giving them the confidence to question an authoritative model. This is evident when we examine how students were reasoning about their data.

We coded students' reasoning into four levels of sophistication, somewhat analogous to Bloom's Taxonomy \cite{Bloom}, with the highest level reached by a student in a given experiment being recorded. Level 1 comments reflect the simple application of analysis tools or comparisons without interpretation; level 2 comments analyze or interpret results; level 3 comments combine multiple ideas or propose something new; and level 4 comments evaluate or defend the new idea (see S1.6 for additional comments and figures S2 and S3 for examples of this coding). 

\begin{figure*}[ht]
\begin{center}
\includegraphics[width=\textwidth]{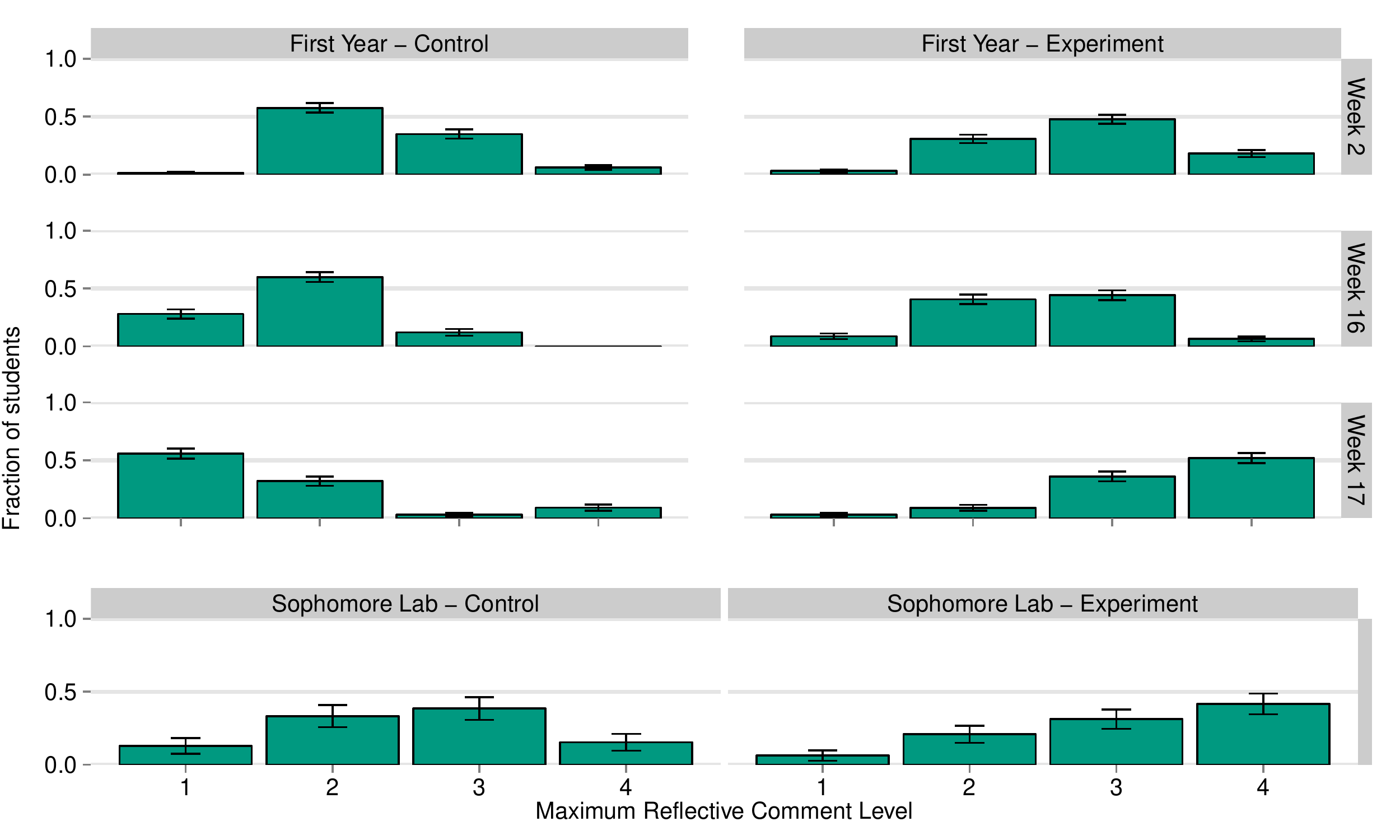}
\caption{\textbf{Reflective comments.} The distribution of the maximum reflection comment level students reached in four different experiments (three in the first year course and one in the sophomore course) shows statistically significant differences between groups (see S2.3 for statistical analyses). Uncertainty bars represent 67\% confidence intervals on the proportions of students.}\label{fig:max}
\end{center}
\end{figure*}

In figure \ref{fig:max}, we see only a moderate difference between the experimental and control groups in week 2, even though the experimental group received significant behavioural support in week 2. This suggests that the support alone is insufficient to create significant behavioural change. By week 16, there is a larger difference between the groups, with the control group shifting to lower levels of comment sophistication and the experimental group maintaining higher levels of comment sophistication, despite the removal of the behavioural support. In week 17, when the model under investigation is inadequate to explain high-quality data, the difference between the groups becomes much more dramatic. For the experimental group, the unexpected disagreement triggers productive, deep analysis of the comparison beyond the level the previous week \cite{PERC2, ProductiveFailure, VanLehn1988}. We attribute this primarily to attempts to correct or interpret the disagreement. In contrast, most of the students in the control group are reduced to simply writing about the analysis tools they had used.

Students in the control group had primarily been analyzing and interpreting results (level one and two), but not acting on them. Since students will continue to use strategies that have been successful in the past \cite{Butler2002}, the students were not prepared to deal with the unexpected outcome in week 17. Our data, however, is limited in that we only evaluate what was written in their books by the end of the lab session. It is plausible that the students in the control group were holding high-level discussions about the disagreement, but not writing them down. Their low-level written reflections are, at best, evidence that they needed more time to achieve the outcomes of the experiment group. 

In the sophomore lab, the students in the experimental group continued to show a high level in their reflective comments, showing a sustained change in reasoning and epistemology. The students in the control group show higher-level reflections in the sophomore lab than they did in the first-year lab, possibly because of the greater time given to analyze their data, the prompt about the model failing, or the selection of these students as physics majors. They still remained well below the level of the experimental group, nonetheless.

\section*{Discussion}
The cycles of making and deciding how to act on quantitative comparisons gave students experience with making authentic scientific decisions about data and models. Since students had to ultimately decide how to proceed, the cycles provided a constrained experimental design space to prepare them for autonomous decision-making \cite{Sere2002}. With a focus on the quality of their data and how they could improve it, they came to believe that they are able to test and evaluate models. This is not just an acquisition of skills; it is an attitudinal and epistemological shift unseen in the control group or in other studies of instructional labs \cite{PERC, Sere2001}. The training in how to think like an expert inherently teaches students how experts think and, thus, how experts generate knowledge \cite{Baron1993}.

The simple nature of the structure employed here gives students both a framework and a habit of mind that leaves them better prepared to transfer the skills and behaviours to new contexts \cite{Bulu2010, SalomonPerkins, Sternberg}. This simplicity also makes it easily generalizable to a very wide range of instructional settings; any venue that contains opportunities to make decisions based on comparisons.

\section*{Acknowledgments}
The authors would like to acknowledge the support of Deborah Butler in preparing the manuscript. We would also like to thank Jim Carolan for the diagnostic survey data about the study participants. This research was supported by UBC's Carl Wieman Science Education Initiative.

\nocite{ErrorBars}
\nocite{GUM}
\nocite{FCI}
\nocite{MBT}
\nocite{BEMA}
\nocite{R}
 \nocite{lme4} 
 \nocite{car}
 \nocite{HofsteinLunetta}

\bibliographystyle{plain}

\end{document}


\title{Teaching critical thinking}

\author{N.G. Holmes}
\email[]{ngholmes@stanford.edu}
\affiliation{Department of Physics, Stanford University, Stanford, CA}
\author{Carl E. Wieman}
\affiliation{Department of Physics, Stanford University, Stanford, CA}
\affiliation{Graduate School of Education, Stanford University, Stanford, CA}
\author{D.A. Bonn}
\affiliation{Department of Physics and Astronomy, University of British Columbia, Vancouver, BC}

\maketitle

\setcounter{figure}{0}
\renewcommand{\theequation}{S\arabic{equation}}
\renewcommand{\thefigure}{S\arabic{figure}}
\renewcommand{\thetable}{S\arabic{table}}
\renewcommand{\thesection}{S\arabic{section}}
\renewcommand{\thesubsection}{S\Alph{subsection}}

\section*{Supplementary Materials}

\section{Quantitative Comparison Tools} 
The first type of comparison encountered in a typical introductory physics lab is to compare two independently measured values of the same physical parameter, a task that is known to be challenging for students [3,5,10]
. In many instructional labs, students do so by assessing whether the uncertainty ranges defined by the measurements overlap. Scientists, however, generally refer to a continuous scale associated with the measurements' probability distributions [22]
, such as the number of units of uncertainty by which two measurements differ (so called $1-\sigma$, $2-\sigma$, or $3-\sigma$ differences in physics, for example). Following the Guide to Uncertainty in Measurement [23]
, this could be calculated as, 

\begin{equation}
t^\prime = \frac{A-B}{\sqrt{\delta_A^2 + \delta_B^2}},
\end{equation}

where $A$ and $B$ are two measured values and $\delta_A$ and $\delta_B$ are their uncertainties, respectively. As such, a large $t^\prime$-score means the measurements differ by more than their combined uncertainties and a small $t^\prime$-score means the measurements are similar within their combined uncertainties. We use the letter $t$ for the index in reference to the structural similarity to the \emph{Student's t}-value, but we do not imply the index applies to the $t$-distribution.

Interpreting the outcome of this comparison provides the necessary structure for deciding how to act on the comparison. For example, since overestimated uncertainties can lead to an artificially small $t^\prime$-score, a low $t^\prime$-score could mean that poor precision has hidden a small disagreement. As such, one could choose to improve the quality of the measurements. Under a model that predicts the two measurements should agree, a large $t^\prime$-score could mean that the model is limited or inappropriate. One could then choose to evaluate, adjust, or discard this model. One could also attempt to identify possible measurement errors that are causing a systematic effect. In all of these cases, the statistic compares the difference between measured quantities within units of variability. Rather than specifically comparing sample means according to the sample standard deviations, however, the $t^\prime$-score uses any measurement value with its uncertainty. As such, we do not try to compare the $t^\prime$-scores on the $t$-distribution or make inferences about probabilities. Indeed, if the measurements were sample means from populations with the same variance, the $t^\prime$-score would be equivalent to \emph{Student's t} for comparing independent samples (or, if homogeneity of variance is violated, the $t^\prime$-score would be equivalent to \emph{Welch's t}).

As discussed in the main text, the $\chi^2$ equation for least squares fitting lends itself to the same quantitative framework defined by the weighted or reduced $\chi^2$ statistic,

\begin{equation}
\chi_w^2 = \frac{1}{N} \sum_{i=1}^{N} {\left( \frac{y_i - f(x_i)}{\delta y_i}\right)^2},
\end{equation}

where $x_i$ and $y_i$ are the measured independent and dependent values, $\delta y_i$ is the uncertainty associated with each $y_i$, $N$ is the number of data points, and $f(x_i)$ are the model values associated with each $x_i$. This parameter evaluates the average difference between measured data and a model in units of uncertainty (squared). Values, therefore, are subject to the same interpretation and follow-up measurements as with the $t^\prime$-score (see table \ref{tab:tab}).

\begin{table*}[ht]
\caption{Interpretations of and follow-up behaviours from $t^\prime$-score comparisons between two measurements or $\chi^2$ comparisons between data sets and models.}
\begin{tabular}{lp{5cm}p{6.5cm}l}
\hline
\hline
$t^\prime$-score & Interpretation of measurements & Follow-up investigation & $\chi^2$-value\\
\hline
$0< |t^\prime|<1$ & Unlikely different, uncertainty may be overestimated & Improve measurements, reduce uncertainty & $0 < \chi^2 < 1$ \\
$1< |t^\prime|<3$ & Unclear whether different & Improve measurements, reduce uncertainty & $1< \chi^2 < 9$ \\
$3< |t^\prime|$ & Likely different & Improve measurements, correct systematic errors, evaluate model limitations or approximations & $9< \chi^2$ \\
\hline
\hline
\end{tabular}
\label{tab:tab}
\end{table*}

Students were also taught a number of additional statistical analysis tools. The full set of tools taught to each condition are found in table \ref{tab:tools}, which also specifies whether the tool informs a comparison or is primarily procedural.

\begin{table*}[ht]
\caption{Statistical tools taught to students in each condition, specified by whether it is procedural or informs the comparison cycles.}
\begin{tabular}{p{1.5cm}|c|c}
\hline
\hline
\multicolumn{2}{c}{Comparison Tools} & Procedural Tools\\
\hline
Control & Experiment & Control \& Experiment\\
\hline
& $t^\prime$-score & Histograms\\
& Residual plots  & Mean\\
\multicolumn{2}{c|}{Overlapping uncertainty ranges}  & Standard deviation \\
\multicolumn{2}{c|}{Unweighted $\chi^2$} & Standard uncertainty in the mean (standard error)\\
\multicolumn{2}{c|}{Weighted $\chi^2$}& Semi-log and log-log plots\\
\multicolumn{2}{c|}{} & Weighted average \\
\multicolumn{2}{c|}{} & Uncertainty in fit parameters of fit lines \\
\hline
\hline
\end{tabular}
\label{tab:tools}
\end{table*}

\section{Comparison cycles instruction across the year}
As mentioned in the main text, students in the experimental group were given explicit instructions to make comparisons between their measurements and/or models and iterate to improve their measurements. These behaviours were also graded and present in a grading rubric.  This support was faded across the course. The explicit instructions in the text were the first to be removed, followed by assigned marks, and eventually instructor support was also removed. A map of this fading process across the year is included in table \ref{tab:scaff}.

\begin{table*}[ht]
\caption{The experimental group received explicit support to make and act on comparisons. The support came in the form of explicit instructions and/or reference in the marking scheme and was faded over time. In the table, an X means that the behavior (comparing or iterating) was supported that week. }
\begin{tabular}{llccccccccccccccccccc}
\hline
\hline
& \multicolumn{19}{c}{Week}\\
&& 1 & 2 & 3 & 4 & 5 & 6 & 7 & 8 & 9 & 10 & 11 & 12 &13 & 14 & 15 & 16 & 17 & 18 & 19 \\
\hline
\multirow{2}{*}{Compare} & Instructions & & X & X & X & X & X & & & & & X & X & &&&&&&\\
& Marking & & X & X & X & X & X & X & & & X& X& X& & X& & X& &&\\
\multirow{2}{*}{Iterate} & Instructions & & X & && X& X& &&&&&&&&&&&&\\
& Marking & & X & && X & X & X & &&&&&&&&&&&\\
\hline
\hline
\end{tabular}
\label{tab:scaff}
\end{table*}

\section{Student experiments included in the study}
\subsection{Week 2: Period of a pendulum as a function of amplitude}
In this experiment, students were asked to measure the period of a pendulum at two (experimental group, $10\degree$ and $20\degree$) or three (control group, $5\degree$, $10\degree$, and $20\degree$) angles of amplitude and compare their measurements. They were not given a model for the process, but most of the students believed from previous experience (high school or college-level physics class) that the period was independent of angle according to the equation:
\begin{equation}
T=2\pi \sqrt{\frac{L}{g}},
\end{equation}
where $L$ is the length of the pendulum, $g$ is the acceleration due to gravity, and $T$ is the period of the pendulum. The derivation of this equation, however, involves an approximation that,
\begin{equation}
\sin\theta \approx \theta,
\end{equation}
for small angles, $\theta$. High precision measurements, therefore, expose this approximation and reveal the difference in the periods at different amplitudes from the second-order correction to this approximation. 

\subsection{Week 16: RC circuit 2}
In this experiment, students studied the voltage decay across a resistor in a parallel Resistor-Capacitor (RC) circuit. This was the second experiment with this equipment and circuit. They measured the time constant ($\tau$) of the voltage decay across the resistor as a function of resistance of the resistor, which is given by the model,
\begin{equation}
\tau=RC.
\end{equation}
In addition to verifying that the relationship between $\tau$ and $R$ was in fact linear with an intercept through the origin, they could compare the capacitance of the capacitor with the value of the slope from a graph of $\tau$ versus $R$. Resistance from other parts of the circuit were negligible in this experiment.

\subsection{Week 17: LR circuit} 
Using a similar measurement procedure to the week 16 experiment, students studied the time constant of the voltage decay ($\tau$) across a resistor in a series Inductor-Resistor (LR) circuit, which is given by the model, 
\begin{equation}
\tau=\frac{L}{R}.
\end{equation}

For this model, the time constant as a function of resistance, plotted as $\frac{1}{\tau}$ versus resistance, would give a straight line with an intercept through the origin. Resistance in the additional components in the circuit, however, is non-negligible here, resulting in a non-zero intercept in the plot. Students could choose whether to perform a one-parameter ($y=mx$) or two-parameter ($y=mx+b$) linear fit to their data, which would cause them to confront the issue of the intercept. Students did not know the inductance of the inductor and so could not make a comparison to the value from the fit. They could check their circuit for a finite (non-infinite) time constant with the resistor set to zero resistance.

\subsection{Sophomore Lab: LRC circuit}
In the LRC circuit experiment, an inductor (L), resistor (R), and capacitor (C) are connected in series, and the equation governing the voltage decay across the resistor is, 
\begin{equation}
\frac{V_R}{V_0} =\frac{1}{\sqrt(1+((\omega^2+\omega_0^2)/(\gamma \omega))^2 )},
\end{equation}

where $V_R$ is the voltage across the resistor, $V_0$ is the amplitude of the input AC voltage source, $\omega$ is the angular frequency of the voltage source, $\omega_0$ is the resonant frequency, and $\gamma$ is the bandwidth. Students fit their data of $\frac{V_R}{V_0}$ as a function of frequency, $\omega$, to determine the parameters $\omega_0$ and $\gamma$. Additional resistance in the circuit beyond the resistance in the resistor, however, means that the ratio of $V_R$ to $V_0$ will never be exactly 1, and so it is necessary to add a third scaling factor, A, to the model, such that,
\begin{equation}
\frac{V_R}{V_o} =\frac{A}{\sqrt(1+((\omega^2+\omega_o^2)/(\gamma \omega))^2 )}.
\end{equation}
Students also measured the parameters $\omega_0$ and $\gamma$ through another experiment and could calculate their values (using measurements of the components R, L, and C) through the definition of these parameters. As such, they had multiple comparisons to make to inform the quality of the fit beyond the analysis of the fit itself. 

\section{Inter-rater reliability}
For all of the data presented, one rater coded all items and another rater coded approximately 10\% of the items. The primary coder was never blind to condition due to the nature of the student products. In the control group, students printed their analysis work from spreadsheets and pasted them into their lab notes, whereas the experimental group submitted their spreadsheets electronically. The second rater, however, was given copies that made him blind to condition.

Inter-rater reliability analysis using Cohen's $\kappa$ statistic was performed to evaluate consistency between raters. Values greater than 0.6 were considered substantial agreement and so do not suggest a need for blind coding. For the quality of reflective comments, the inter-rater reliability for the raters was found to be $\kappa = 0.657, p<.001$. For identifying whether students proposed or proposed and carried out changes to their methods and measurements, the inter-rater reliability for the raters was found to be $\kappa = 0.714, p<.001$. For identifying whether students identified and/or physically interpreted the disagreements with models, the inter-rater reliability for the raters was found to be $\kappa = 0.881, p<.001$.

\section{Participants}
Included in the study were two cohorts (groups) of students enrolled in the same introductory undergraduate physics course at a research-intensive university in Canada. The control group consisted of students enrolled in 2012/2013, while the experiment group consisted of students enrolled in 2013/2014. The course, both years, was spread across two semesters of eight or nine 3-hour lab weekly lab sessions. Each lab session included no more than 48 students and was facilitated by two graduate student teaching assistants and the course instructor. The number of students included in the analysis is found in table \ref{tab:N}. The variability in the number of students each week is due to students not attending all labs. In the control group, 109 students conducted all three first-year labs and only 31 students conducted all three first-year labs and the sophomore lab. In the experiment group, 108 students conducted all three first-year labs and only 36 students conducted all three first-year labs and the sophomore lab. Since the effects of the lab occurred throughout more than just the four labs evaluated, we include any students who participated each particular week.

\begin{table}[h]
\caption{Sample sizes on each measure in the study between groups and experiments.}
\begin{tabular}{lcccc}
\hline
\hline
Group & Week 2 & Week 16 & Week 17 & Sophomore Lab\\
\hline
Control & 146 & 132 & 131 &39\\
Experiment & 159 & 138 & 133 & 48\\
\hline
\hline
\end{tabular}
\label{tab:N}
\end{table}

On entering the course, the two groups had statistically equivalent pre-test scores on the Force Concept Inventory [24]
: Control, $M=77\%, SE=2\%$; Experiment, $M=76\%, SE=2\%, t(266)=0.20, p=0.839$. By the end of the first term, the groups had statistically equivalent scores on the Mechanics Baseline Test [25]
: Control, $M=72\%, SE=2\%$; Experiment, $M=68\%, SE=2\%, t(288)=1.21, p=0.227$. By the end of the second term, the groups also had statistically equivalent scores on the Brief Electricity and Magnetism Survey [26]
: Control, $M=70\%, SE=2\%$; Experiment, $M=64\%, SE=2\%, t(177)=1.96, p=0.052$. These assessments have been used to evaluate the introductory physics students in the department for over 20 years and, in the last decade, students' incoming scores have been consistent within a 2\% standard deviation.

The critical thinking behaviours assessed in this study relate primarily to evaluating data and physical measurement systems. The questions on the MBT and BEMA evaluate students' ability to apply specific physics concepts in idealized situations. There is very little overlap between the knowledge and reasoning required to answer those questions, and the real-world, data-driven critical thinking about data and measurement systems learned in the lab course. We also would expect that the lecture and other components of the courses would dominant over a possible effect related to the lab. Therefore, it is not surprising that the scores are not correlated. 

Students in the course both years were almost all intending to major in a science, technology, engineering, or math field, though they do not declare their majors until their second year. The breakdown of students' intended majors in the experiment group by the end of the course are in table \ref{tab:majors}. Unfortunately, these data were unavailable for the control group. We do have data that shows that approximately 15\% of students in the control group and 20\% of the students in the experimental group chose physics as a major by their second year of study.

\begin{table}[h]
\caption{The percentage of students in the experimental group who have declared a variety of STEM majors.}
\begin{tabular}{lcccc}
\hline
\hline
Intended Major & Percentage of the experimental group\\
\hline
Physics or Astronomy & 14\%\\
Life Sciences & 13\%\\
Engineering Physics & 7\%\\
Non-STEM & 2\%\\
Computer Science & 1\%\\
Chemistry &  1\%\\
Other STEM or undecided &  62\%\\
\hline
\hline
\end{tabular}
\label{tab:majors}
\end{table}

\subsection{Evaluation of the sophomore students}

We will further evaluate the students who continued into the sophomore lab course to explore whether the results seen in the sophomore lab are due to transfer or selection effects. First, we will do a 2-by-2 comparison on the end-of-first year MBT and BEMA scores (Table \ref{tab:SophMBT}), comparing between students who did and did not take the sophomore lab course and between the experiment and control groups in the first-year course.

Overall, the students who went on to take the sophomore physics lab course outperformed the students who did not take the sophomore lab, as measured on both the MBT and the BEMA (note that, of the students in the control group, there was no difference between students who did and did not take the sophomore lab course on the BEMA). This tells us that the students in the sophomore physics labs generally had a stronger conceptual physics background than the students who did not continue in an upper-year physics lab course. This is consistent with the expected selection bias of students who choose to pursue more physics courses. Of the students who took the sophomore physics lab, however, there is a non-significant difference between the experimental and control groups on both the MBT and BEMA. This is consistent with the overall lack of differences on these measures between the full experiment and control conditions in the first-year lab course discussed in the previous section. 

\begin{table}[h]
\caption{Comparing the students who went into the sophomore physics lab with the students who did not in each cohort on the MBT and BEMA diagnostics at the end of first year.}
\begin{tabular}{lcccc}
MBT & &\\
\hline
\hline
Group & \multicolumn{2}{c}{Sophomore Lab}\\
& Took Lab & Did not take Lab \\
\hline
Control Group & 77(12) & 70(16)\\
Experimental Group & 75(17) & 66(16)\\
\hline
\multicolumn{3}{c}{Comparisons}\\
Control Group & Took lab vs did not take lab & $t$(76.6)=2.46, $p$=.016$^*$\\
Experimental Group & Took lab vs did not take lab & $t$(80.6)=2.81, $p$=.006$^{**}$\\
Took lab & Experimental vs Control Group & $t$(71.2)=0.59, $p$=.556\\
\hline
\hline
&&\\
BEMA & & \\
\hline
\hline
Group & \multicolumn{2}{c}{Sophomore Lab}\\
& Took Lab & Did not take Lab \\
\hline
Control Group & 74(9) & 65(20)\\
Experimental Group & 68(16) & 61(16)\\
\hline
\multicolumn{3}{c}{Comparisons}\\
Control Group & Took lab vs did not take lab & $t$(34.8)=1.85, $p$=.073 \\
Experimental Group & Took lab vs did not take lab & $t$(70.8)=2.06, $p$=.04$^{*}$\\
Took lab & Experimental vs Control Group & $t$(44.3)=1.71, $p$=.094 \\
\hline
\hline
\end{tabular}
\label{tab:SophMBT}
\end{table}

\begin{figure*}
\begin{center}
\begin{subfigure}[b]{.65\textwidth}
\centerline{\includegraphics[width=\textwidth]{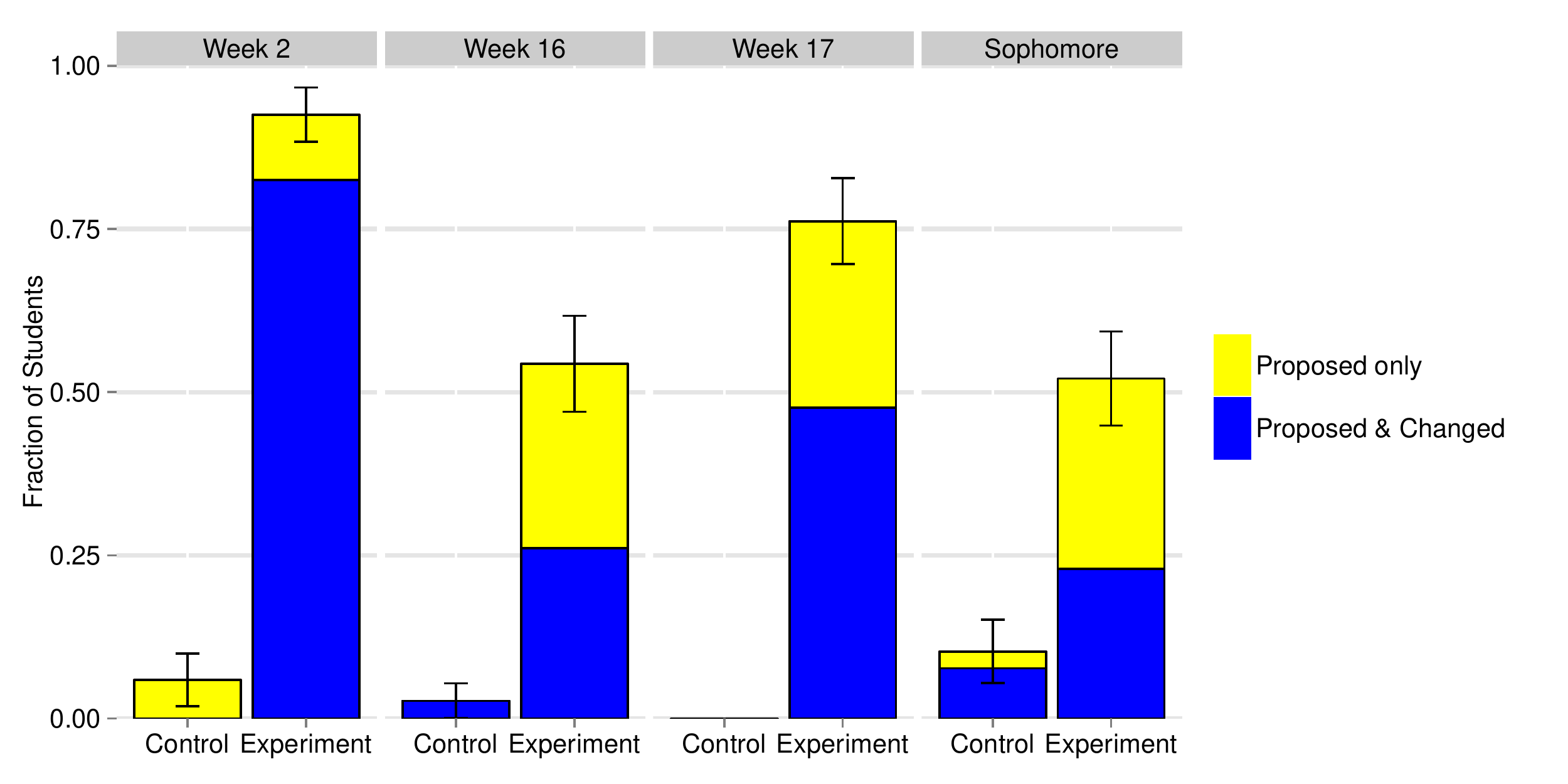}}
\caption{Proposing and/or carrying out changes to their experimental methods}\label{fig:iterateSoph}
\end{subfigure}
\begin{subfigure}[b]{.65\textwidth}
\centerline{\includegraphics[width=\textwidth]{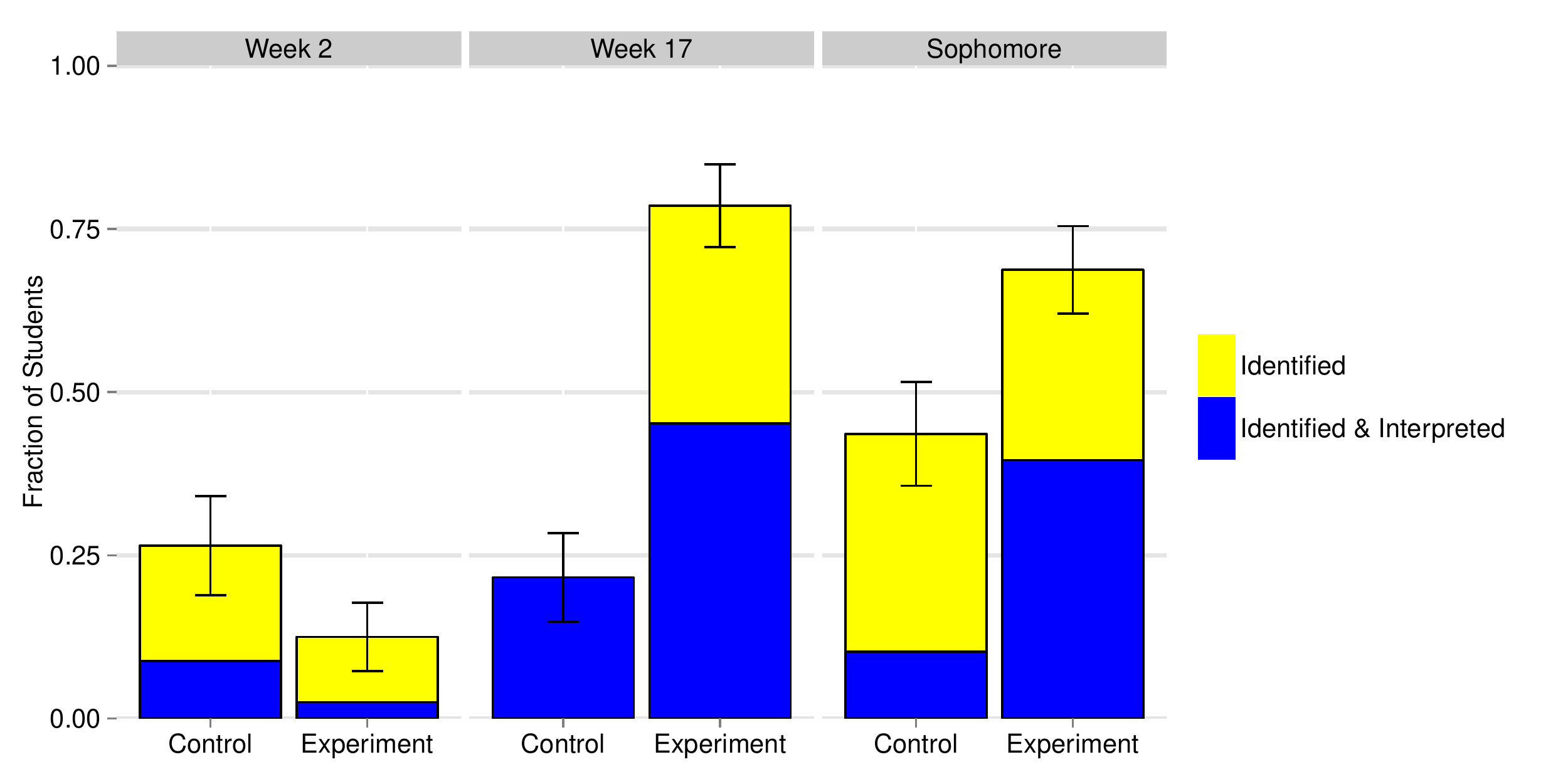}}
\caption{Identifying and interpreting model disagreements}\label{fig:evalSoph}
\end{subfigure}

\begin{subfigure}[b]{.65\textwidth}
\centerline{\includegraphics[width=\textwidth]{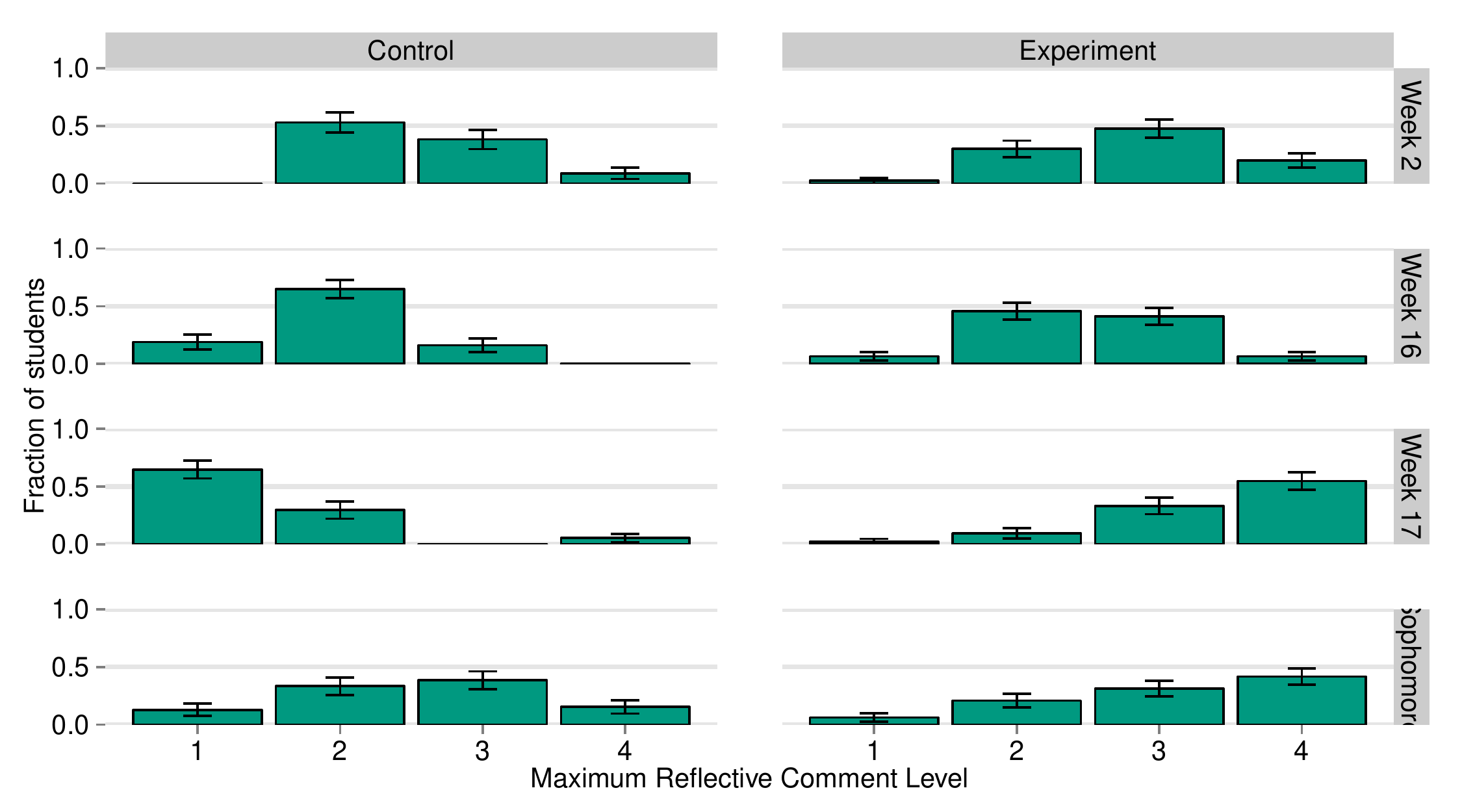}}
\caption{Maximum reflection comment level reached}\label{fig:maxSoph}
\end{subfigure}
\caption{\textbf{Evaluating the sophomore lab students.} The figures show the measures in the core of the analysis including only the students who moved into the sophomore-level physics course. The data is very similar to the class as a whole, demonstrating that the students in the sophomore lab are representative of the full first-year class on these measures.}\label{fig:Soph}
\end{center}
\end{figure*}

Next, we compare these two subgroups on their evaluation, iteration, and reflection behaviours throughout the first year labs. The trends in the figures \ref{fig:evalSoph}, \ref{fig:iterateSoph}, and \ref{fig:maxSoph} showing only the sophomore students are very similar to those for the whole course (figures 1, 2, and 3
). This suggests that the students who continued into the the sophomore course were not exceptional in their behaviours in first-year. This further suggests that the effect seen in the Sophomore Lab experiment are not due to selection effects. It remains that the upwards shift in the control group's reflective comments and evaluation of the model are due to something inherent in the sophomore lab course. Most likely these shifts can be attributed to the prompt in the instructions to explain why there may be extra parameters in the model. This instruction would explain a shift in the model evaluation and reflective comments, but not in iteration, as seen in the data.

\section{Reflection analysis}
\begin{figure*}[h]
\begin{center}
\begin{subfigure}{0.5\textwidth}
\centerline{\includegraphics[width=\textwidth]{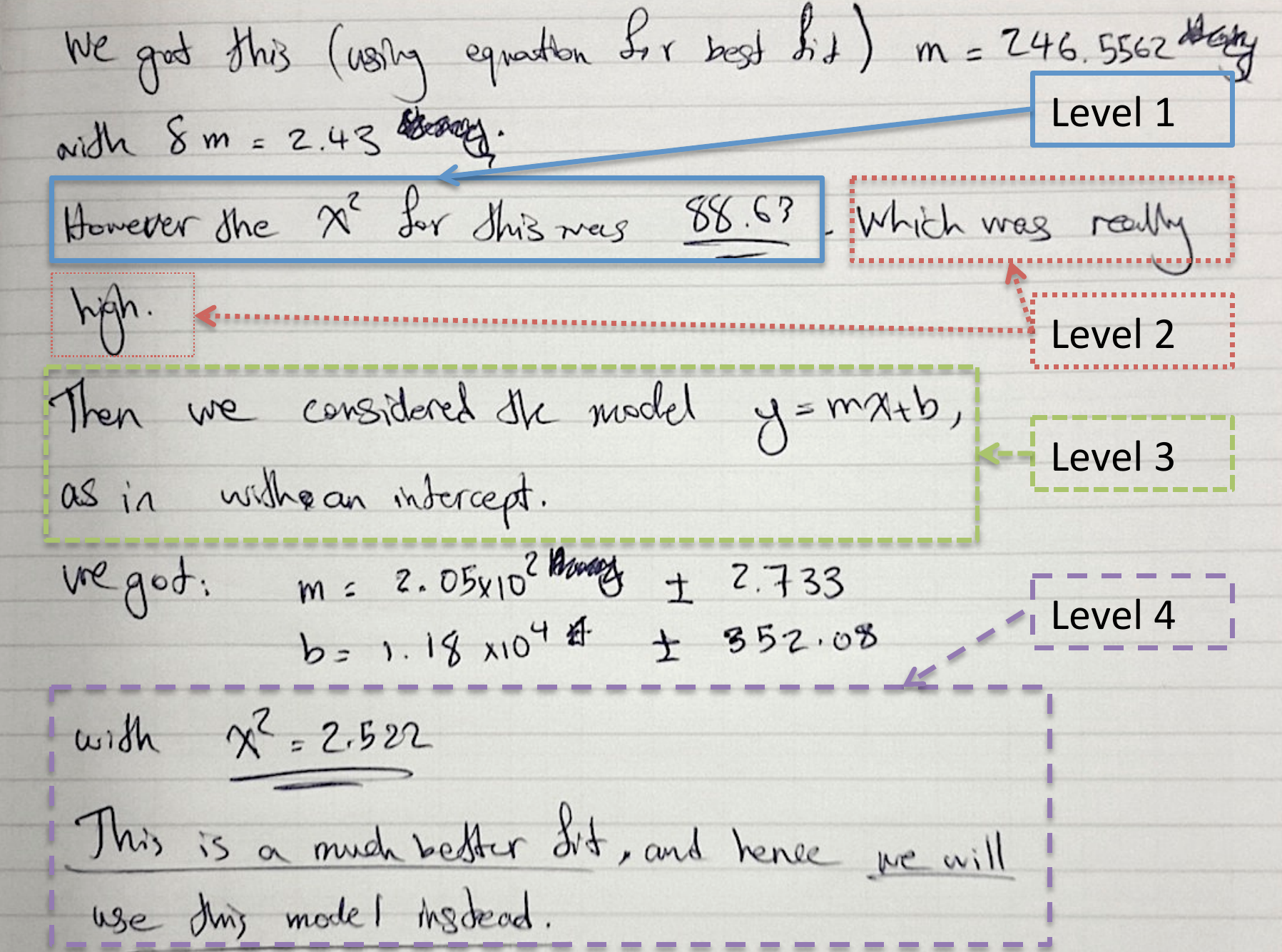}}
\caption{}\label{fig:leva}
\end{subfigure}

\begin{subfigure}{0.5\textwidth}
\centerline{\includegraphics[width=\textwidth]{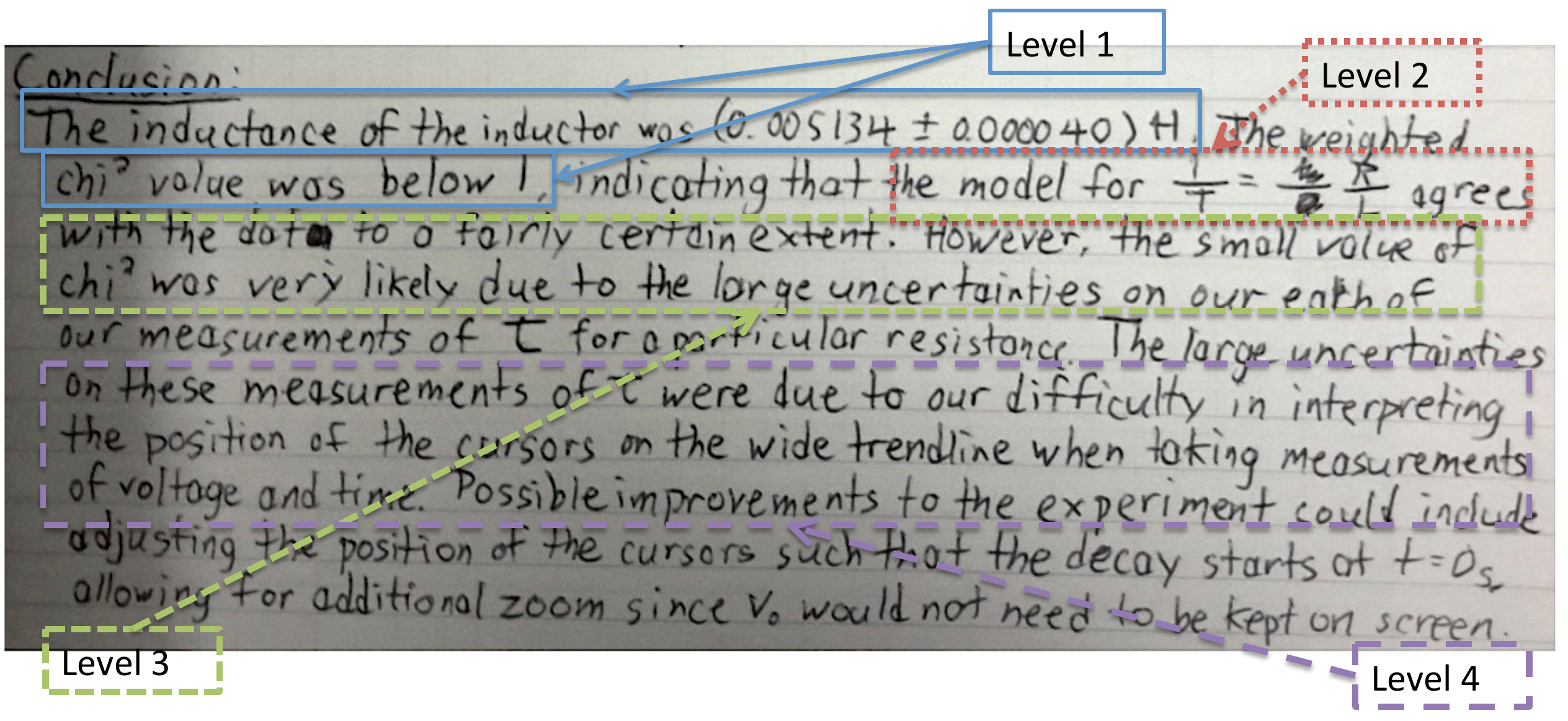}}
\caption{}\label{fig:levb}
\end{subfigure}
\caption{Two students' reflections during an experiment provide examples of the reflection coding scheme. a) The student makes a level 1 comment about applying $\chi^2$ to their experiment, then shows that this value is high (level 2). A level 3 statement describes considering a different model, and then the student finally evaluates the new model by describing the much lower $\chi^2$ value. b) The students starts with a level 1 comment about $\chi^2$ and the inductance, then analyzes the fit line compared to the model (level 2). They then comment on $\chi^2$ being small, attributing it to large uncertainties (level 3). They justify their uncertainty due to limitations of the measurement equipment (level 4). Finally they provide further suggestions for improvement (additional level 3).}\label{fig:Levels}
\end{center}
\end{figure*}

To analyze students' reflection in the lab, we evaluated students' reflective comments associated with their statistical data analysis and conclusions. The reflective comments were coded using a set of four classes based on Bloom's Taxonomy classes [13]
. Figures \ref{fig:leva} and \ref{fig:levb} provide samples of this coding applied to student work. The four comments levels were:
\begin{enumerate}
\item Application - a written reflection statement that offers the outcome of the procedural application of data analysis tools (e.g. The $\chi^2$ value is 2.1.) These comments were distinct from procedural statements (e.g. Then we calculated the $\chi^2$ value.)
\item Analysis - a written reflection statement that analyzes or interprets their data analysis or results (e.g. Our $\chi^2$ value is 0.84, which is close to one, indicating that our model fits the data well.)
\item Synthesis - a written reflection statement that synthesizes multiple ideas, tool analyses, or reflections to propose a new idea. This could include suggesting ways to improve measurements (e.g. we will take more data in this range, since the data is sparse) or models (e.g. our data has an intercept so the model should have an intercept), as well as making comparisons (e.g. The $\chi^2$ value for the $y=mx$ fit was 43.8, but for the $y=mx+b$ fit $\chi^2$ was 4.17, which is much smaller.)
\item Evaluation - a written reflection statement that evaluates, criticizes, or judges the previous ideas presented. Evaluation can look similar to analysis, but the distinction is that evaluation must follow a synthesis comment. For example, after a synthesis that compared two different models and demonstrated that adding an intercept lowered the $\chi^2$ value, an evaluation could follow as, ``...the intercept was necessary due, most likely, to the inherent resistance within the circuit (such as in the wires)."
\end{enumerate}

Figures \ref{fig:leva} and \ref{fig:levb} demonstrate how the coding scheme is applied to three excerpts from students' books in the LR experiment (week 17). Each of the levels build on each other, so a student making a level 4 evaluation statement would also have made lower level statements, though level 1 comments (application) need not be present. While it is important that students reflect on various parts of the data analysis, the results presented in the main text examine the maximum reflection level a student reached. It should be noted that the comments were not evaluated on correctness. 

\section{Analysis}
For the first-year experiments, generalized linear mixed-effects models were performed using R [27] 
 and and the Linear Mixed-Effects Models using `Eigen' and S4 package [28] 
to analyze all three outcome measures (proposing and/or carrying out measurement changes, identifying and/or interpreting disagreements with models, and levels of reflection/comments). For measurement changes and evaluating models, logistic regression analysis was performed due to the dichotomous nature of the outcome variables. For the reflection data, Poisson regression was used due to account for the bounded nature of the outcome variables. All three analyses used condition, lab week, and the interaction between condition and lab week as fixed effects and Subject ID as a random effects intercept. Type 3 analysis of variance (ANOVA) was performed on the logistic regression models using the R Companion to Applied Regression package [29] 
to assess the overall impact of the variables. Sophomore lab data were analyzed using $\chi^2$ tests for independence of proportions.

\subsection{Proposing and/or carrying out measurement changes}
A logistic regression was carried out to compare the proportion of students in each group and across each experiment proposing and/or carrying out changes to their measurements (table \ref{tab:Anova1}). Note, for this analysis, proposing versus proposing and carrying out changes were collapsed to a single dichotomous variable of proposing or carrying out changes. The logistic regression model was statistically significant, $\chi^2(5)=470.55, p<.001$. A Type 3 ANOVA of the logistic regression model demonstrated that condition and the interaction between condition and lab week were highly significant in the model, but lab week alone was not significant. 

\begin{table*}[ht]
\caption{Results from the logistic regression comparing students' iteration behaviours in each group across four experiments.}
\begin{tabular}{lrrrr@{.}l}
\hline
\hline
Model coefficients & Estimate & $S.E.$ & Wald $z$ &\multicolumn{2}{c}{$p$}\\
\hline
Condition = Experiment & 7.97 & 0.94 & 8.49 & $<$&$0001^{***}$\\
Week = Week 16 & -0.82 & 0.86 & -0.96 & &336\\
Week = Week 17 & -0.41 & 0.75 & -0.55 & &582\\
$[Condition = Experiment] * [Week = Week 16]$& -2.64 & 1.03 & -2.56 & &$010^{**}$\\
$[Condition = Experiment] * [Week = Week 17]$& -2.54 & 0.93 & -2.72 & &$007^{**}$\\
\hline
Model variables & &$df$ & $\chi^2$ & $p$ &\\
\hline
Condition & &1 & 83.02 & $<$&$001^{***}$\\
Week & &2 & 28.99 & $<$&$001^{***}$\\
Condition$*$Week & & 2 & 9.28 & &$01^*$\\
\hline
\multicolumn{6}{l}{$^*p<.05, ^{**}p<.01, ^{***}p<.001$.}\\
\hline
\hline
\end{tabular}
\label{tab:Anova1}
\end{table*}

With significant effects for the interaction, we can compare the groups each week to explore where the significant differences exist. To do this, we use a $\chi^2$ test of proportions comparing groups on the distribution of the number of students who did not propose or change their measurements, who proposed changes to their measurements, and who proposed and made changes to their measurements (returning to the three-level, rather than dichotomous, variable). Taking into account the multiple comparisons across weeks, we use a Bonferroni correct to set the $\alpha$-level at .01. This gave statistically significant differences between groups on all four experiments: Week 2, $\chi^2(2)=270.38, p<.001$; Week 16, $\chi^2(2)=107.51, p<.001$; Week 17, $\chi^2(2)=128.39, p<.001$; Sophomore Lab, $\chi^2(2)=17.58, p<.001$. This demonstrates that the experiment group outperformed the control group on this measure on all experiments.

\subsection{Evaluating models}
A logistic regression was carried out to compare the proportion of students in each group and across each experiment identifying the disagreement with the model and/or physically interpreting the issue (table \ref{tab:Anova2}). Note, for this analysis, identifying versus physically interpreting the disagreement with the model were collapsed to a single dichotomous variable. The logistic regression model was statistically significant, $\chi^2(3)=171.96, p<.001$. A Type 3 ANOVA of the logistic regression model demonstrated that condition and the interaction between condition and lab week were highly significant in the model, but lab week alone was not significant. 

\begin{table*}[ht]
\caption{Results from the logistic regression comparing students' behaviours identifying disagreements with a given model (ID) and physically interpreting the disagreement (PI) across four experiments.}
\begin{tabular}{lrrrr@{.}l}
\hline
\hline
Model coefficients & Estimate & $S.E.$ & Wald $z$ & \multicolumn{2}{c}{$p$}\\
\hline
Condition = Experiment & -0.83 & 0.33 & -2.55 & &$011^{*}$\\
Week = Week 17 & -0.27 & 0.30 & -0.88 & &379\\
$[Condition = Experiment] * [Week = Week 17]$& 3.60 & 0.60 & 5.97 & $<$&$001^{***}$\\
\hline
Model variables & &$df$ & $\chi^2$ & $p$ & \\
\hline
Condition & &1 & 6.49 & & $011^{*}$\\
Week & &1 & 0.77 & &379\\
Condition$*$Week & &1 & 35.62 & $<$&$001^{***}$\\
\hline
\multicolumn{6}{l}{$^*p<.05, ^{**}p<.01, ^{***}p<.001$.}\\
\hline
\hline
\end{tabular}
\label{tab:Anova2}
\end{table*}

With significant effects for the interaction, we can compare the groups each week to explore where the significant differences exist. To do this, we use a $\chi^2$ test of proportions comparing groups on the distribution of the number of students who did not identify the disagreement with a model, who did identify the disagreement, and who identified and interpreted the disagreement. Taking into account the multiple comparisons across weeks, we use a Bonferroni correct to set the $\alpha$-level at .02. This gave significant differences between groups on all three experiments: Week 2, $\chi^2(2)=8.60, p=.014$; Week 17, $\chi^2(2)=99.04, p<.001$; Sophomore Lab, $\chi^2(2)=10.32, p=.006$.

\subsection{Reflection behaviours}
A Poisson regression was carried out to analyze the quality of the reflective comments in each group across each experiment (table \ref{tab:Anova3}). The regression model was statistically significant, $\chi^2(5)=109.03, p<.001$. A Type 3 ANOVA of the logistic regression model demonstrated that condition and the interaction between condition and lab week were highly significant in the model, but lab week alone was not significant. 

\begin{table*}[ht]
\caption{Results from the regression comparing students' maximum reflection level in each group across four experiments.}
\begin{tabular}{lrrrr@{.}l}
\hline
\hline
Model coefficients & Estimate & $S.E.$ & Wald $z$ &  \multicolumn{2}{c}{$p$}\\
\hline
Condition = Experiment & 0.13 & 0.07 & 1.89 & &$059 .$\\
Week = Week 16 & -0.29 & 0.08 & -3.48 &$<$ &$001^{***}$\\
Week = Week 17 & -0.40 & 0.09 & -4.59 & $<$&$001^{***}$\\
$[Condition = Experiment] * [Week = Week 16]$& 0.17 & 0.11 & 1.52 & &130\\
$[Condition = Experiment] * [Week = Week 17]$& 0.58 & 0.11 & 5.29 & $<$&$001^{***}$\\
\hline
Model variables && $df$ & $\chi^2$ & $p$ & \\
\hline
Condition & &1 & 3.57 & &$059 .$\\
Week & &2 & 24.48 & $<$&$001^{***}$\\
Condition$*$Week && 2 & 28.55 & $<$&$001^{***}$\\
\hline
\multicolumn{6}{l}{$. p<.1, ^*p<.05, ^{**}p<.01, ^{***}p<.001$.}\\
\hline
\hline
\end{tabular}
\label{tab:Anova3}
\end{table*}

With a significant interaction, we can compare the groups each week to explore where the significant differences exist. To do this, we use a $\chi^2$ test of proportions comparing the distribution of the numbers of students in each group who reached each maximum comment level. Taking into account the multiple comparisons across weeks, we use a Bonferroni correct to set the $\alpha$-level at .01. This gave significant differences between groups on all three first-year experiments, but non-significant differences on the sophomore-lab: Week 2, $\chi^2(3)=25.44, p<.001$; Week 16, $\chi^2(3)=51.86, p<.0001$; Week 17, $\chi^2(3)=155.83, p<.0001$; Sophomore Lab, $\chi^2(3)=7.58, p=.056$.

\section{Time on task in the LR experiment}

One confounding issue to the week 17 LR circuit experiment was that students in the control group worked through a computer-based inquiry activity at the beginning of the experiment session. The activity taught students how to calculate the uncertainty in the slope of a best-fitting line, which they also used to reanalyze the previous week's data. As such, the control group spent approximately two hours on the LR circuit lab, whereas the experiment group spent three hours. Not having enough time to reflect on data and act on that reflection may explain the different outcomes observed in the main text. As a precautionary measure, we observed students in the experiment group two-hours into the lab session to evaluate what analysis they had performed by that time. The observer recorded whether the group had by that time produced a one-parameter $mx$ fit or a two-parameter $mx+b$ fit.

\begin{figure*}[ht]
\begin{center}
\centerline{\includegraphics[width=.5\textwidth]{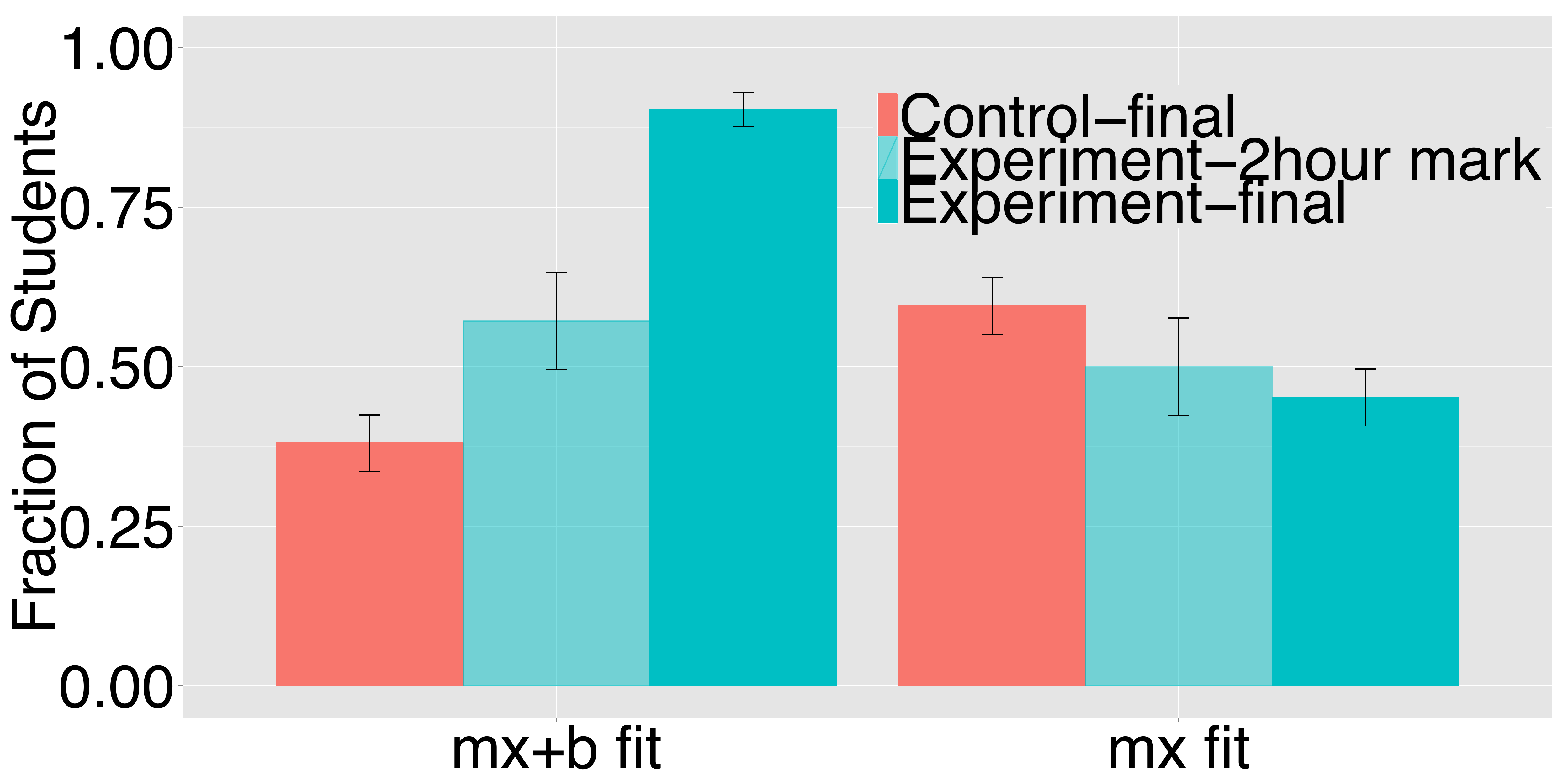}}
\caption{The distribution of graphical analyses made by students by the end of the LR circuits lab in the control and experiment groups and within the first two-hours of the lab for the experiment group. Uncertainty bars represent 67\% confidence intervals on the proportions. They are larger for the ``Experiment-2hour mark," since only groups, rather than individuals, were assessed. Bars in each group may add to more than 1, since students may have analyzed either or both fits.}\label{fig:tot}
\end{center}
\end{figure*}

The results, shown in figure \ref{fig:tot}, demonstrate that if the students in the experiment group had been given the same amount of time on task as students in the control group, more of them still would have made the modification to the model and included an intercept in their fit. Given additional time, however, even more students were able to think critically about the task and make better sense of their data. From this result, we conclude that the effects seen in this experiment are still primarily due to students' overall improved behaviours. Indeed, the effect is much larger due to the additional time, which is an important feature of the intervention itself. It takes time for students to engage deeply in a task, think critically, and solve any problems that arise [30]
. Comparing between students in the experiment group at the 2-hour mark and the final 3-hour mark demonstrates the striking effect that an extra hour can make to students' productivity.

The number of single-parameter $mx$ fits decreased slightly from the 2-hour observations and the final submitted materials for the experiment group. This could have occurred if students recognized that the $mx$ fit was not helpful in understanding their data, due to the additional intercept required. This is interesting to note in light of the limitations of the analysis methods used in this study. Analyzing lab books can only keep track of recorded activity and many behaviours may have occurred without record. The result that some students created additional fits and then did not submit them at the end of the lab period demonstrates that students in the experiment group still may have engaged in additional reflective and iterative behaviours beyond what was recorded. Differences between the control and experiment groups, then, are unlikely attributed to students in the experiment group simply recording more while engaging in the same behaviours as students in the control group.

The slope uncertainty activity provided to the students in the control group just before the LR circuit lab may, however, have narrowed the focus of students' analysis. That is, the activity first introduced students to the uncertainty in the slope of a one-parameter best fitting line (that is, with the intercept fixed at the origin). As such, it could be argued that these students were more likely to fix the intercept at the origin so that they could apply the learned formula. The activity, however, also included a follow-up task that introduced the uncertainty in the slope of a two-parameter best fitting line (intercept not fixed) and so students did have access to both options. They also could have used their analysis to identify the issue even if they did not change their fit.
%
